\documentclass[10pt,twocolumn,superscriptaddress,amssymb,amsmath,aps,pra]{revtex4-1}

\usepackage{graphicx}

\begin{document}


\title{Time-dependent optical force theory for optomechanics of dispersive 3D photonic materials and devices}
\date{July 20, 2022}
\author{Mikko Partanen}
\affiliation{Photonics Group, Department of Electronics and Nanoengineering, Aalto University, P.O. Box 13500, 00076 Aalto, Finland}
\author{Jukka Tulkki}
\affiliation{Engineered Nanosystems Group, School of Science, Aalto University, P.O. Box 12200, 00076 Aalto, Finland}

\begin{abstract}
We present a position- and time-dependent optical force theory for optomechanics of dispersive 3D photonic materials and devices. The theory applies to media including material interfaces, waveguides, and general photonic crystal structures. The theory enables calculation of the dynamical state of the coupled field-material system and the interference of this state with other excitations of the material, such as surface acoustic waves or phonons. As an example, we present computer simulations of energy and momentum flows through a silicon crystal with anti-reflective structured interfaces. Using commercially available simulation tools, the theory can be applied to analyze optical forces in complex photonic materials and devices.
\end{abstract}

\maketitle

\onecolumngrid
\vspace{-0.2cm}
\twocolumngrid

\section{Introduction}

The electric and magnetic fields of light in arbitrary 3D photonic materials and devices can be unambiguously solved from Maxwell's equations. Considering the interest to develop optomechanic materials and devices, it is astonishing that no generally applicable \emph{time-dependent} optical force theory has been presented for \emph{inhomogeneous dispersive materials} \cite{Sun2015,Rakich2010}. The optical force density is ultimately related to the conservation law of momentum. Optical forces having the same physical origin as those discussed in the present work have already led to important scientific discoveries and photonic technologies in optical trapping and laser cooling. Optical forces are also related to the centenary Abraham-Minkowski controversy of the difference of the momenta of light in a material and in vacuum \cite{Leonhardt2006a,Barnett2010b,Kemp2011,Milonni2010,Bliokh2017a}. In previous literature, most works assume the Maxwell or Helmholtz stress tensor and further take the \emph{time average} of the resulting optical force density \cite{Sun2015,Wang2016a,Wang2016b,Rakich2010,Rakich2011,Brevik2018c,Pernice2009}. Very few works, if any, are studying the \emph{exact position- and time-dependent} optical force density in inhomogeneous dispersive materials. However, the knowledge of the exact position- and time-dependence of the optical force density is of fundamental importance for \emph{quantitative analysis} of the interaction of light with elastic waves or other mechanical eigenmodes of the material.

The recently introduced mass-polariton (MP) theory of light \cite{Partanen2021b,Partanen2017c,Partanen2017e,Partanen2019a,Partanen2019b,Partanen2018b} considers a light pulse propagating in a material as a coupled state of the electromagnetic (EM) field and the material and splits the total momentum of light into the field and material components. The material component of the momentum is carried by an atomic mass density wave (MDW) driven forward by optical forces. In this work, we generalize the MP theory for \emph{structured interfaces} and give a general expression for the optical force density in an \emph{inhomogeneous dispersive} dielectric. The optical force density is derived from the continuity equations of the energy and momentum at interfaces and from the covariance principle of the special theory of relativity. The optical force density is tested by computer simulations of energy and momentum flows through a silicon crystal with anti-reflective structured interfaces \cite{Deinega2011}.

The exact time-dependent force density enables accurate calculation of the time- and position-dependent dynamics of the material at structured interfaces and in the bulk, which can also be a \emph{photonic crystal}. This provides an interesting possibility to discover eventual coupling of the field-driven material dynamics in photonic materials and devices with acoustic waves and a possibility to fine tune this coupling for improved or new operational characteristics of photonic devices.

This paper is organized as follows. In Sec.~\ref{sec:separation}, we describe the physical foundations for the separation of the energy density, momentum density, and the stress tensor of the field-material system unambiguously into the field and material parts. In Sec.~\ref{sec:conservation}, the differential forms of the conservation laws of energy and momentum are briefly presented for the field and the material. In Sec.~\ref{sec:restframe}, we solve for the quantities appearing in the conservation laws in the special case of the laboratory frame, i.e., in the rest frame of the material.  In Sec.~\ref{sec:simulation}, we present an example simulation of a Gaussian plane wave light pulse propagating through a silicon crystal block with anti-reflective structured interfaces. In Sec.~\ref{sec:SEM}, we present the relativistically invariant expressions of the stress-energy-momentum (SEM) tensors of the field and material parts of the system. Finally, conclusions are drawn in Sec.~\ref{sec:conclusions}.

\section{\label{sec:separation}Separation of the system into field and material parts}

\subsection{Separation of the energy density, momentum density, and the stress tensor}

To enable unambiguous separation of the energy density, momentum density, and the stress tensor of the field-material system into the field and material parts, we define the energy density of the material to include only the classical \emph{rest energy density and kinetic energy density} of atoms. All other forms of the energy density, such as the energy density of the polarization field, are considered to be field energy density. It is well known that under the influence of classical EM field, the atomic trajectories follow very accurately classical Newtonian dynamics, and therefore, their rest energy and kinetic energy are strictly unambiguous and equal to their classical values. Therefore, \emph{within our definition} of the energy density of the material, there cannot be any uncertainties in the splitting of the total energy density or momentum density into the field and material parts. This means that the energy and momentum densities and the stress tensor of the material in a general inertial frame are uniquely given by the classical formulas $W_\mathrm{mat}=\rho_\mathrm{a}c^2$, $\mathbf{G}_\mathrm{mat}=\rho_\mathrm{a}\mathbf{v}_\mathrm{a}$, $\boldsymbol{\mathcal{T}}_\mathrm{mat}=\rho_\mathrm{a}\mathbf{v}_\mathrm{a}\otimes\mathbf{v}_\mathrm{a}$, respectively, where $\rho_\mathrm{a}$ is the mass density, $\mathbf{v}_\mathrm{a}$ is the atomic velocity, $c$ is the speed of light in vacuum, and $\otimes$ denotes the outer product of vectors \cite{Dirac1996,Misner1973}. The mass density is given by $\rho_\mathrm{a}=\gamma_{\mathbf{v}_\mathrm{a}}m_0n_\mathrm{a}$, where $m_0$ is the rest mass of an atom, $n_\mathrm{a}$ is the number density of atoms, and $\gamma_{\mathbf{v}_\mathrm{a}}=1/\sqrt{1-|\mathbf{v}_\mathrm{a}|^2/c^2}$ is the Lorentz factor. The relations above preserve Newton's equation of motion of the material, in all inertial frames, in the form $n_\mathrm{a}\frac{d}{dt}\mathbf{p}_\mathrm{a}=\mathbf{f}_\mathrm{opt}$, where $\mathbf{p}_\mathrm{a}=\gamma_{\mathbf{v}_\mathrm{a}}m_0\mathbf{v}_\mathrm{a}$ is the momentum of an atom and $\mathbf{f}_\mathrm{opt}$ is the optical force density \cite{Penfield1967,Partanen2021b}. The number density of atoms satisfies the continuity equation $\frac{\partial}{\partial t}n_\mathrm{a}+\nabla\cdot(n_\mathrm{a}\mathbf{v}_\mathrm{a})=0$ \cite{Penfield1967,Partanen2021b}. Note that the elastic force density and the electro- and magnetostrictive force densities between atoms, discussed in Sec.~\ref{sec:fst}, can, however, modify the energy and momentum densities and the stress tensor of the material above as briefly discussed in Sec.~\ref{sec:SEM}. The energy density, momentum density, and the stress tensor of the field in a general inertial frame are denoted by $W_\mathrm{EM}$, $\mathbf{G}_\mathrm{EM}$, and $\boldsymbol{\mathcal{T}}_\mathrm{EM}$, respectively.

Note that all quantities in the general inertial frame are \emph{unambiguously determined by their values in the laboratory frame} by their respective well-known Lorentz transformations. The Lorentz transformations needed are collected in Table V of Ref.~\cite{Partanen2021b}, which also presents a complete record of the covariant theory of light in dispersive materials. In spite of using a different expression of the optical force and the SEM tensors, in this work, the covariance properties presented in Ref.~\cite{Partanen2021b} are \emph{as is} strictly applicable and fulfilled for the theory presented. Since, in our theory, the energy density, momentum density, and the stress tensor of the material are uniquely determined by the classical formulas as explained above, the separations of the total energy density, momentum density, and the stress tensor of the field-material system into the field and material parts are also unique. In previous literature, some works claim that the separation of the momentum density of light into the field and material parts would be arbitrary \cite{Pfeifer2007}, while some other works base their separation of the system into the field and material parts on arguments that are different from those of us \cite{Sheppard2016a,Sheppard2016b}. The previous challenges in the unambiguous splitting of the system into the field and material parts may be related to adding to the energy density of the material, for instance, the energy density of the polarization field, which has its origin in the field energy.

\subsection{\label{sec:fst}Separation of the total field-induced force density into gradient forces and forces that carry the wave momentum of light}

In addition to the optical force density $\mathbf{f}_\mathrm{opt}$ acting between the propagating EM field and the material atoms along the wave vector of light, there are optically induced intra-material force densities, called electro- and magnetostriction, which appear in the form of a gradient force $\mathbf{f}_\mathrm{st}$ acting along optical energy density gradients \cite{Landau1984,Gordon1973,Washimi1976,Stratton1941}. Very recently, the electrostrictive force density has also been verified experimentally for an optical field \cite{Astrath2022}. The splitting of the total force density into gradient forces and forces associated with the wave propagation has also been discussed in a recent preprint \cite{Bliokh2022a}. The optical electro- and magnetostrictive force densities $\mathbf{f}_\mathrm{st}$ rise from the dependence of the optical electric and magnetic energy densities on the atomic density through the permittivity and permeability of the material \cite{Landau1984}. The integral of the gradient force densities over the volume of the material is zero at any time, in particular when a light pulse is entering the material. Thus, these force densities do not lead to the movement of the center of mass of the material even though they give rise to local accelerations of the material atoms. Therefore, $\mathbf{f}_\mathrm{st}$ differs from $\mathbf{f}_\mathrm{opt}$, which is a topic of the present work. In the present work, we have preserved the term optical force to mean only forces that carry the wave momentum of light. Consequently, we focus on the investigation of the dynamical effects of $\mathbf{f}_\mathrm{opt}$ and leave the detailed time-dependent study of $\mathbf{f}_\mathrm{st}$ in dispersive materials as a topic of a future work.

\section{\label{sec:conservation}Conservation laws of energy and momentum}

We solve the optical force density in an inhomogeneous dispersive material starting from the conservation laws of energy and momentum for the EM field, written for a \emph{general inertial frame} in a differential form as \cite{Landau1984,Jackson1999,Partanen2021b,Penfield1967,Philbin2011}
\begin{equation}
 \frac{1}{c^2}\frac{\partial W_\mathrm{EM}}{\partial t}+\nabla\cdot\mathbf{G}_\mathrm{EM}=-\frac{\phi_\mathrm{opt}}{c^2},
 \label{eq:energyconservation}
\end{equation}
\begin{equation}
 \frac{\partial\mathbf{G}_\mathrm{EM}}{\partial t}+\nabla\cdot\boldsymbol{\mathcal{T}}_\mathrm{EM}=-\mathbf{f}_\mathrm{opt}.
 \label{eq:momentumconservation}
\end{equation}
Here $\phi_\mathrm{opt}=\mathbf{f}_\mathrm{opt}\cdot\mathbf{v}_\mathrm{a}$ is the optical power-conversion density of a lossless material. In the \emph{rest frame} of lossless materials, i.e., the laboratory frame, the optical power-conversion density can be approximated to zero as $\phi_\mathrm{opt}^\mathrm{(L)}=0$, since it only converts an exceedingly small amount of EM energy to the kinetic energy of atoms \cite{Partanen2021b}. The conservation laws for the energy and momentum of the material are, accordingly, given for a general inertial frame in a differential form by \cite{Partanen2021b}
\begin{equation}
 \frac{1}{c^2}\frac{\partial W_\mathrm{mat}}{\partial t}+\nabla\cdot\mathbf{G}_\mathrm{mat}=\frac{\phi_\mathrm{opt}}{c^2},
 \label{eq:energyconservation2}
\end{equation}
\begin{equation}
 \frac{\partial\mathbf{G}_\mathrm{mat}}{\partial t}+\nabla\cdot\boldsymbol{\mathcal{T}}_\mathrm{mat}=\mathbf{f}_\mathrm{opt}.
 \label{eq:momentumconservation2}
\end{equation}
Here it is essential to note that the source terms on the right hand side are opposite to those of the EM field in the conservation laws in Eqs.~\eqref{eq:energyconservation} and \eqref{eq:momentumconservation}. Thus, the total quantities of the field-material system, i.e., sums of the field and material parts, satisfy the conservation laws in the form, where the source terms are zero. When the conservation laws in Eqs.~\eqref{eq:energyconservation}--\eqref{eq:momentumconservation2} are satisfied, it is automatically guaranteed that the energy and momentum are conserved at material interfaces and that the law of constant center of energy velocity is satisfied for an isolated system.

\section{\label{sec:restframe}Energy, momentum, and optical force densities in the laboratory frame}

\subsection{\label{sec:energyandmomentumdensities}Energy and momentum densities of the field}

The EM contribution to the total momentum density of light is continuous over material interfaces and given in the laboratory frame by \cite{Landau1984,Jackson1999}
\begin{equation}
 \mathbf{G}_\mathrm{EM}^\mathrm{(L)}=\frac{\mathbf{E}^\mathrm{(L)}\times\mathbf{H}^\mathrm{(L)}}{c^2}.
 \label{eq:GEM}
\end{equation}
Here $\mathbf{E}^\mathrm{(L)}$ is the electric field and $\mathbf{H}^\mathrm{(L)}$ is the magnetic field solved in the laboratory frame from Maxwell's equations with position- and frequency-dependent permittivity and permeability. Using the EM momentum density in Eq.~\eqref{eq:GEM}, the exact \emph{time-dependent} energy density of the EM field can be solved from the conservation law of energy in Eq.~\eqref{eq:energyconservation} with $\phi_\mathrm{opt}^\mathrm{(L)}=0$. For a \emph{narrow frequency band} in a locally isotropic lossless dispersive material, where the permittivity and permeability are real-valued scalars, the EM energy density becomes \cite{Partanen2021b}
\begin{align}
 W_\mathrm{EM}^\mathrm{(L)} &=\frac{1}{2}\Big(\varepsilon|\mathbf{E}^\mathrm{(L)}|^2+\mu|\mathbf{H}^\mathrm{(L)}|^2\nonumber\\
 &\hspace{0.4cm}+\omega_0\frac{\partial\varepsilon}{\partial\omega_0}\langle|\mathbf{E}^\mathrm{(L)}|^2\rangle+\omega_0\frac{\partial\mu}{\partial\omega_0}\langle|\mathbf{H}^\mathrm{(L)}|^2\rangle\Big).
 \label{eq:WEM}
\end{align}
Here $\omega_0$ is the central angular frequency of the narrow frequency band, and $\varepsilon=\varepsilon(\omega_0)$ and $\mu=\mu(\omega_0)$ are the permittivity and permeability of the material in the laboratory frame at $\omega_0$. For a narrow frequency band, the permittivity can be approximated by the first two terms of its Taylor series $\varepsilon(\omega)\approx\varepsilon(\omega_0)+\frac{\partial\varepsilon(\omega_0)}{\partial\omega_0}(\omega-\omega_0)$. A similar expansion is valid for the permeability. The angle brackets in Eq.~\eqref{eq:WEM} are defined by $\langle|\mathbf{E}^\mathrm{(L)}|^2\rangle=\frac{1}{2}[|\mathbf{E}^\mathrm{(L)}|^2+|\frac{1}{\omega_0}\frac{\partial}{\partial t}\mathbf{E}^\mathrm{(L)}|^2]$ and $\langle|\mathbf{H}^\mathrm{(L)}|^2\rangle=\frac{1}{2}[|\mathbf{H}^\mathrm{(L)}|^2+|\frac{1}{\omega_0}\frac{\partial}{\partial t}\mathbf{H}^\mathrm{(L)}|^2]$. The brackets here and below are also equal to the time average of the pertinent quantities over the harmonic cycle, e.g., $\langle|\mathbf{E}^\mathrm{(L)}(\mathbf{r},t)|^2\rangle=\frac{1}{T}\int_{-T/2}^{T/2}|\mathbf{E}^\mathrm{(L)}(\mathbf{r},t+t')|^2dt'$, where $T=2\pi/\omega_0$ is the length of the harmonic cycle. The EM energy density in Eq.~\eqref{eq:WEM} is equal to that in Ref.~\cite{Partanen2021b}, but it is expressed here in terms of the permittivity and permeability of the material and their derivatives. In previous literature, there are some works that add to the EM momentum density of the field in Eq.~\eqref{eq:GEM} additional dispersion related terms  \cite{Veselago1968,Kemp2007}. In contrast, in the present work, we rely on the EM momentum density in Eq.~\eqref{eq:GEM} and show that it leads to a consistent theory. In our theory, dispersion dependence appears in the momentum density of the EM field only in the general inertial frame as presented in Sec.~\ref{sec:SEM}.

\subsection{\label{sec:f}Optical force density and stress tensor of the field}

Having the energy density of the EM field solved from the conservation law of energy in Eq.~\eqref{eq:energyconservation}, we are left with a problem of determining the EM stress tensor and the optical force density from the conservation law of momentum, given in Eq.~\eqref{eq:momentumconservation}. \emph{Appriori}, if we are only given the conservation law of momentum in Eq.~\eqref{eq:momentumconservation} with the EM momentum density in Eq.~\eqref{eq:GEM}, both the EM stress tensor and the optical force density are unknown. Therefore, more information is needed to determine these quantities and to guarantee their uniqueness. In Ref.~\cite{Partanen2021b}, we \emph{postulated} for the optical force density, in the rest frame of a homogeneous material, a generalized expression of the Abraham force, given by $\mathbf{f}_\mathrm{opt,A}^\mathrm{(L)}=\frac{n_\mathrm{p}n_\mathrm{g}-1}{c^2}\frac{\partial}{\partial t}(\mathbf{E}^\mathrm{(L)}\times\mathbf{H}^\mathrm{(L)})$, where $n_\mathrm{p}$ and $n_\mathrm{g}$ are the phase and group refractive indices, respectively, defined in the laboratory frame. Then, we used $\mathbf{f}_\mathrm{opt,A}^\mathrm{(L)}$ to determine the corresponding EM stress tensor \cite{Partanen2021b}. In Ref.~\cite{Partanen2021b}, it was proven that the force density $\mathbf{f}_\mathrm{opt,A}^\mathrm{(L)}$ leads to a consistent covariant theory. However, it was also pointed out that the covariance condition and the continuity equation are not sufficient to quarantee the uniqueness of the optical force density. In this work, we use a different approach by introducing an additional constraint to the available EM energy density and the total momentum density of light. This constraint is shown to lead to unambiguous expressions for the optical force density and the EM stress tensor.

To obtain the necessary constraint, we argue that, for a light pulse propagating in a homogeneous material, for every differential volume element, the total momentum density of light must have the same constant proportionality to the EM energy density. Without this local property, if an arbitrarily small amount of energy is absorbed from the EM field, the absorbed momentum would depend on the space-time point, where the absorption takes place. This would imply that the energy-momentum ratio of the non-absorbed part of the light pulse also changes, which is not physically meaningful. The constant of proportionality is required to be $n_\mathrm{p}/c$, which agrees with the high-precision experimental results for the radiation pressure in dispersive liquids by Jones \emph{et al.} \cite{Jones1978}. Thus, the total momentum of light $\mathbf{G}_\mathrm{EM}^\mathrm{(L)}+\mathbf{G}_\mathrm{mat}^\mathrm{(L)}$ is required to satisfy $\mathbf{G}_\mathrm{EM}^\mathrm{(L)}+\mathbf{G}_\mathrm{mat}^\mathrm{(L)}=n_\mathrm{p}W_\mathrm{EM}^\mathrm{(L)}\hat{\mathbf{v}}_\mathrm{g}^\mathrm{(L)}/c$, where $\hat{\mathbf{v}}_\mathrm{g}$ is a unit vector in the direction of the group velocity of light. Using Newton's equation of motion, we can write $\mathbf{G}_\mathrm{mat}^\mathrm{(L)}=\rho_\mathrm{a}^\mathrm{(L)}\mathbf{v}_\mathrm{a}^\mathrm{(L)}=\int_{-\infty}^t\mathbf{f}_\mathrm{opt}^\mathrm{(L)}dt'$, which relates the given condition unambiguously to the expression of the optical force density. After some algebra, the optical force density is given by
\begin{align}
 \mathbf{f}_\mathrm{opt}^\mathrm{(L)}
 &=\frac{\partial}{\partial t}\Big[\frac{n_\mathrm{p}^2-1}{c^2}\mathbf{E}^\mathrm{(L)}\!\!\times\!\mathbf{H}^\mathrm{(L)}\!+\!\frac{n_\mathrm{p}(n_\mathrm{g}-n_\mathrm{p})}{c^2}\langle\mathbf{E}^\mathrm{(L)}\!\!\times\!\mathbf{H}^\mathrm{(L)}\rangle\Big]\nonumber\\
 &\hspace{0.4cm}-\frac{1}{2}|\mathbf{E}^\mathrm{(L)}|^2\nabla\varepsilon-\frac{1}{2}|\mathbf{H}^\mathrm{(L)}|^2\nabla\mu.
 \label{eq:fopt}
\end{align}
Here the angle bracket expression is given by $\langle\mathbf{E}^\mathrm{(L)}\times\mathbf{H}^\mathrm{(L)}\rangle=\frac{1}{2}[\mathbf{E}^\mathrm{(L)}\times\mathbf{H}^\mathrm{(L)}+(\frac{1}{\omega_0}\frac{\partial}{\partial t}\mathbf{E}^\mathrm{(L)})\times(\frac{1}{\omega_0}\frac{\partial}{\partial t}\mathbf{H}^\mathrm{(L)})]$. This time- and position-dependent quantity is also equal to the time average of the Poynting vector over the harmonic cycle. In Eq.~\eqref{eq:fopt}, we have added the last two well-known gradient terms \cite{Landau1984} to conform with the conservation law of momentum in Eq.~\eqref{eq:momentumconservation} at \emph{material interfaces}. The EM stress tensor corresponding to the optical force density in Eq.~\eqref{eq:fopt} is found to be given by
\begin{align}
 \boldsymbol{\mathcal{T}}_\mathrm{EM}^\mathrm{(L)} &=\frac{1}{2}\Big[(\mathbf{E}^\mathrm{(L)}\cdot\mathbf{D}^\mathrm{(L)}+\mathbf{H}^\mathrm{(L)}\cdot\mathbf{B}^\mathrm{(L)})\mathbf{I}-\mathbf{E}^\mathrm{(L)}\otimes\mathbf{D}^\mathrm{(L)}\nonumber\\
 &\hspace{0.4cm}-\mathbf{D}^\mathrm{(L)}\otimes\mathbf{E}^\mathrm{(L)}-\mathbf{H}^\mathrm{(L)}\otimes\mathbf{B}^\mathrm{(L)}-\mathbf{B}^\mathrm{(L)}\otimes\mathbf{H}^\mathrm{(L)}\Big].
 \label{eq:TEM}
\end{align}
Here $\mathbf{I}$ is the $3\times3$ unit matrix. Equation \eqref{eq:TEM} is the symmetrized form of the conventional Abraham stress tensor \cite{Pfeifer2007,Kemp2017,Partanen2021b}. The symmetrization is essential in birefringent materials, where the electric and magnetic flux densities $\mathbf{D}^\mathrm{(L)}$ and $\mathbf{B}^\mathrm{(L)}$ are not parallel to the electric and magnetic fields $\mathbf{E}^\mathrm{(L)}$ and $\mathbf{H}^\mathrm{(L)}$, respectively\cite{Landau1984}. The optical force density in Eq.~\eqref{eq:fopt} and the EM stress tensor in Eq.~\eqref{eq:TEM} lead to a theory that fulfills the same covariance conditions as the theory presented in Ref.~\cite{Partanen2021b}. Accordingly, the integrated value of the momentum density of a homogeneous material over the volume of the light pulse following from $\mathbf{f}_\mathrm{opt}^\mathrm{(L)}$ is equal to that obtained using $\mathbf{f}_\mathrm{opt,A}^\mathrm{(L)}$.

\subsection{\label{sec:MDW}Atomic mass density wave}

We next consider the implications of the optical force density to the dynamical state of the material. We define the energy and momentum densities of the atomic MDW as differences between the energy densities and momentum densities of the material under the influence of the EM field and in the absence of it as $W_\mathrm{MDW}=W_\mathrm{mat}-W_\mathrm{mat,0}=\rho_\mathrm{a}c^2-\rho_\mathrm{a0}c^2$ and $\mathbf{G}_\mathrm{MDW}=\mathbf{G}_\mathrm{mat}-\mathbf{G}_\mathrm{mat,0}=\rho_\mathrm{a}\mathbf{v}_\mathrm{a}-\rho_\mathrm{a0}\mathbf{v}_\mathrm{a0}$. Here $\rho_\mathrm{a0}$ is the atomic mass density in the absence of light and $\mathbf{v}_\mathrm{a0}$ is correspondingly the atomic velocity in the absence of light, i.e., $\mathbf{v}_\mathrm{a0}^\mathrm{(L)}=\mathbf{0}$ in the laboratory frame. The momentum density of the material atoms, $\mathbf{G}_\mathrm{mat}^\mathrm{(L)}=\rho_\mathrm{a}^\mathrm{(L)}\mathbf{v}_\mathrm{a}^\mathrm{(L)}=\int_{-\infty}^t\mathbf{f}_\mathrm{opt}^\mathrm{(L)}dt'$, resulting from the optical force density, is obtained as a solution of Newton's equation of motion. Using the expression of the optical force density in Eq.~\eqref{eq:fopt} then gives the momentum density of the atomic MDW in the laboratory frame as
\begin{equation}
 \mathbf{G}_\mathrm{MDW}^\mathrm{(L)}=(n_\mathrm{p}^2-1)\mathbf{G}_\mathrm{EM}^\mathrm{(L)}+n_\mathrm{p}(n_\mathrm{g}-n_\mathrm{p})\langle\mathbf{G}_\mathrm{EM}^\mathrm{(L)}\rangle.
 \label{eq:GMDW}
\end{equation}
The excess energy density of the material, $W_\mathrm{MDW}^\mathrm{(L)}$, can be solved from the continuity equation of energy of the material, which is equivalent to Eq.~\eqref{eq:energyconservation2} for $\phi_\mathrm{opt}^\mathrm{(L)}=0$. Using Eq.~\eqref{eq:GMDW} and $W_\mathrm{MDW}^\mathrm{(L)}=\rho_\mathrm{MDW}^\mathrm{(L)}c^2=\rho_\mathrm{a}^\mathrm{(L)}c^2-\rho_\mathrm{a0}^\mathrm{(L)}c^2$, we obtain the excess mass density $\rho_\mathrm{MDW}^\mathrm{(L)}=\rho_\mathrm{a}^\mathrm{(L)}-\rho_\mathrm{a0}^\mathrm{(L)}$ in a homogeneous region of space, given by
\begin{equation}
 \rho_\mathrm{MDW}^\mathrm{(L)}=\frac{n_\mathrm{p}^2-1}{c^2}W_\mathrm{EM}^\mathrm{(L)}+\frac{n_\mathrm{p}(n_\mathrm{g}-n_\mathrm{p})}{c^2}\langle W_\mathrm{EM}^\mathrm{(L)}\rangle.
 \label{eq:rhoMDW}
\end{equation}
Other dynamical quantities of the material, such as the strain, are also obtained from the solution of Newton's equation of motion.

\section{\label{sec:simulation}Simulation of force and momentum densities}

\begin{figure}
\centering
\includegraphics[width=0.98\columnwidth]{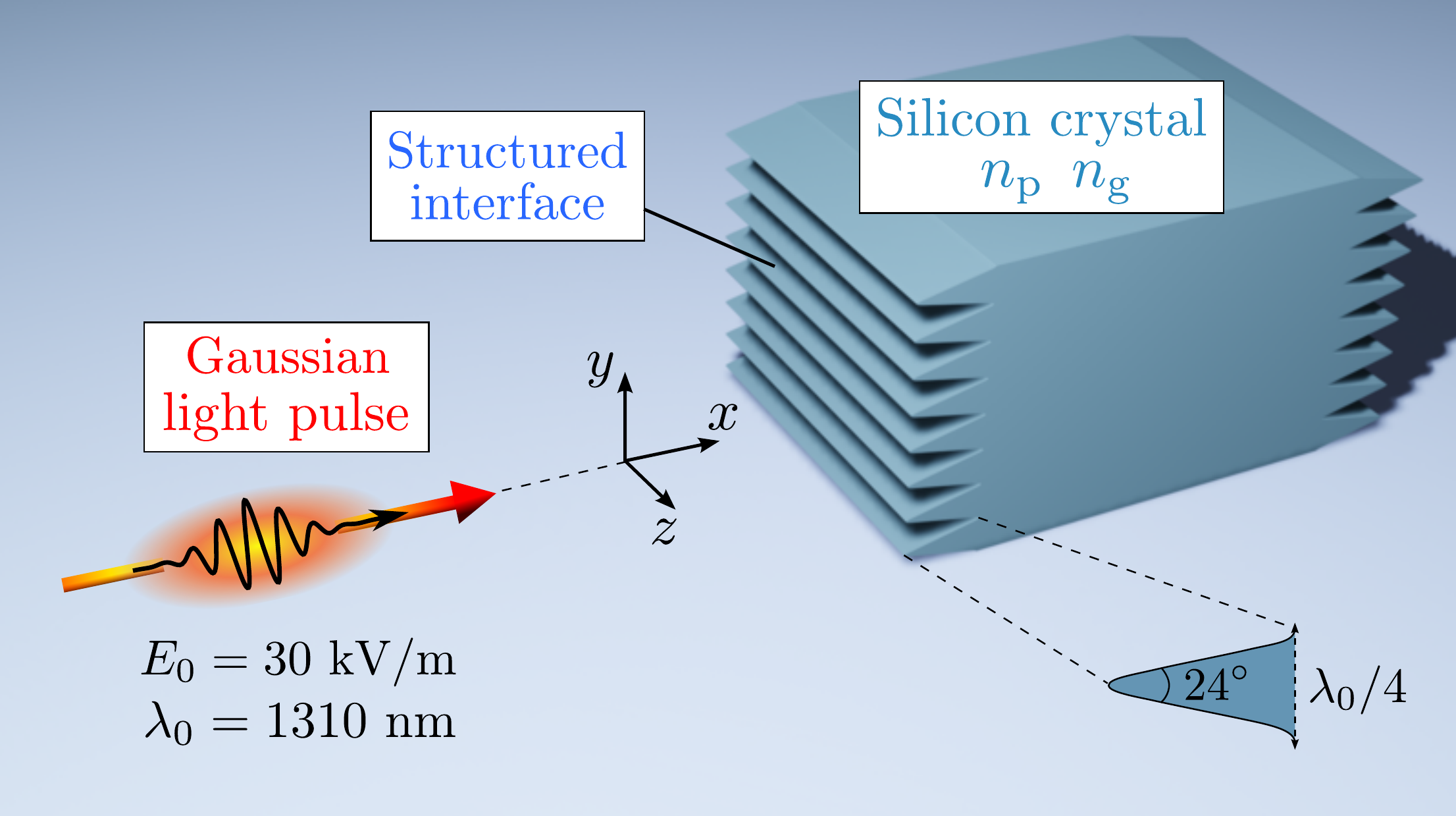}
\caption{\label{fig:illustration}
We have tested our unified optical force theory by simulating the fulfillment of the energy and momentum conservation laws when a light pulse propagates through a silicon crystal block with anti-reflective structured interfaces.}
\end{figure}

To investigate the physical implications of the unified optical force theory presented above, we have simulated light in various geometries including several material interfaces. All of these studies support the expression of the optical force density in Eq.~\eqref{eq:fopt}. As an example, we present numerical simulations of the propagation of a Gaussian plane wave light pulse through a silicon crystal block with anti-reflective structured interfaces illustrated in Fig.~\ref{fig:illustration}. Minimizing reflection with structured interfaces has been discussed more extensively in Ref.~\cite{Deinega2011}.

In the simulations, we assume an incident Gaussian plane wave light pulse with a central vacuum wavelength $\lambda_0=1310$ nm. For this wavelength, the phase refractive index profile of silicon, based on a linear fit to the experimental data from Ref.~\cite{Schinke2015}, is $n_\mathrm{p}=3.5039$ and the group refractive index is $n_\mathrm{g}=3.6840$. The electric field of the Gaussian plane wave light pulse propagating in vacuum in the direction of the positive $x$ axis and polarized along the $y$ axis is given by
\begin{equation}
 \mathbf{E}^\mathrm{(L)}(\mathbf{r},t)=E_0\cos\Big[k_0\Big(x-ct\Big)\Big]
 e^{-(\Delta k_0)^2(x-ct)^2/2}\hat{\mathbf{y}}.
 \label{eq:electricfield}
\end{equation}
Here $k_0=\omega_0/c$ is the wave number in vacuum, $\Delta k_0$ is the standard deviation of the wave number in vacuum, and $E_0$ is the normalization factor corresponding to the peak value of the electric field of the incident Gaussian light pulse. In the simulations, we use $\Delta k_0=0.025k_0$ and $E_0=30$ kV/m. The electric flux density, magnetic field, and magnetic flux density follow unambiguously from the electric field in Eq.~\eqref{eq:electricfield} through Maxwell's equations.  The simulations are carried out using Comsol Multiphysics simulation tool \cite{Comsol2020}. Periodic boundary conditions are applied in the direction of the $y$ axis in Fig.~\ref{fig:illustration}. The periodicity of $\lambda_0/4$ is determined by the periodicity of the structured vacuum-silicon interface.

\begin{figure}
\centering
\includegraphics[width=0.98\columnwidth]{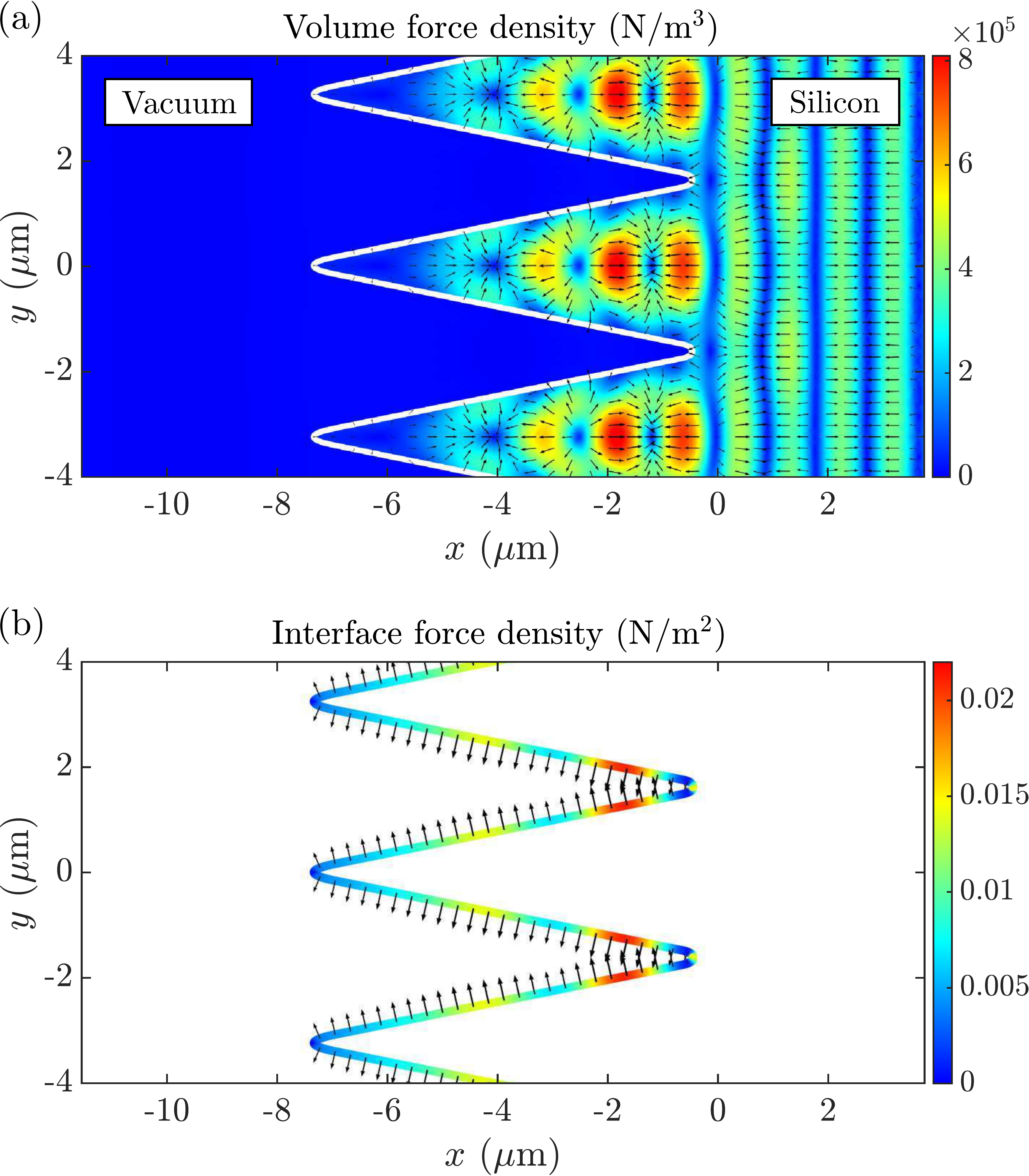}
\caption{\label{fig:forcedensities}
Simulated (a) volume force density and (b) interface force density at an instance of time when the center of the light pulse is at the position of the entry interface of the silicon crystal (see Visualization 1). The light pulse is incident from vacuum on the left. The colors show the magnitudes, and the logarithmically scaled arrows show the directions.}
\vspace{-0.2cm}
\end{figure}

\begin{figure*}
\centering
\includegraphics[width=\textwidth]{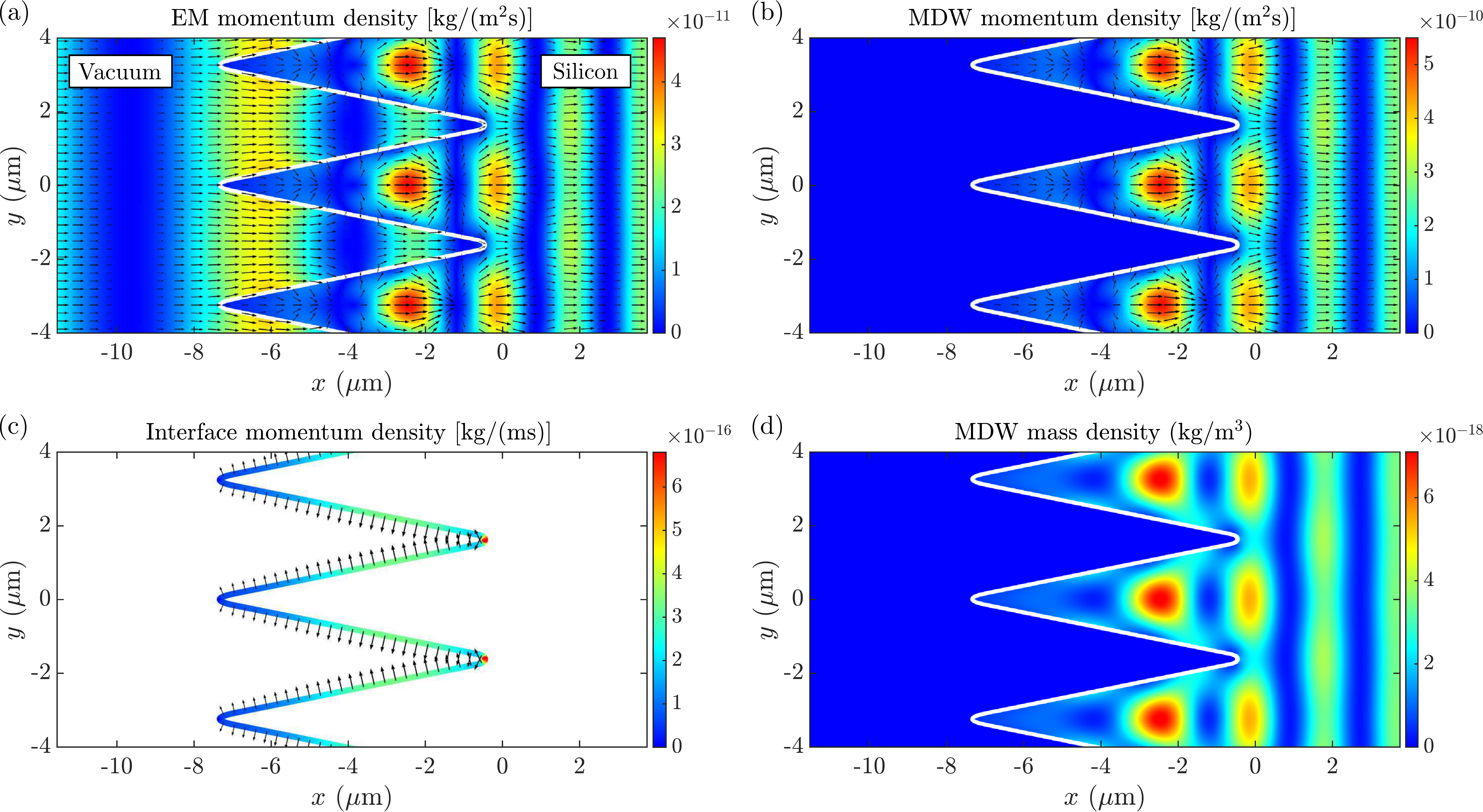}
\vspace{-0.45cm}
\caption{\label{fig:graphs}
Simulated (a) momentum density of the EM field, (b) momentum density of the atomic MDW, (c) momentum density of interface atoms, and (d) mass density of the atomic MDW at an instance of time when the center of the light pulse is at the position of the entry interface of the silicon crystal (see Visualization 2). The light pulse is incident from vacuum on the left. The colors show the magnitudes, and the logarithmically scaled arrows show the directions.}
\vspace{-0.1cm}
\end{figure*}

Figure \ref{fig:forcedensities} presents the optical volume and interface force densities due to the Gaussian light pulse (see Visualization 1). The snapshot is taken at the instance of time when the center of the light pulse is at the position of the entry interface of the silicon crystal.  The volume force density in Fig.~\ref{fig:forcedensities}(a) and the interface force density in Fig.~\ref{fig:forcedensities}(b) have both $x$ and $y$ components that vary as a function of the position. The force density maxima are around positions where the field is focused by the local lensing effect of the vacuum-silicon interface. However, after the interface, the focusing fades out and the force density follows the plane wave form of the field.

Figures \ref{fig:graphs}(a)--(d) show the position dependencies of the EM momentum density, MDW momentum density, interface momentum density, and the MDW mass density, respectively, at the instance of time when the center of the light pulse is at the position of the entry interface of the silicon crystal corresponding to the force densities in Fig.~\ref{fig:forcedensities} (see Visualization 2). The EM momentum density in Fig.~\ref{fig:graphs}(a) consists of the incident and transmitted contributions. The reflection is close to zero due to the anti-reflection property of the structured interface. The atomic MDW momentum density in Fig.~\ref{fig:graphs}(b) reminds the EM momentum density inside silicon on the right and there is no MDW in vacuum on the left. The interface momentum density at all positions in Fig.~\ref{fig:graphs}(c) is directed toward vacuum. Figure \ref{fig:graphs}(d) presents the mass density of the atomic MDW, which is driven forward by the optical volume force density. This transfer of mass and the related rest energy is crucial for the fulfillment of the covariance principle of the special theory of relativity \cite{Partanen2017c,Partanen2017e}.

\begin{figure}
\centering
\includegraphics[width=0.48\textwidth]{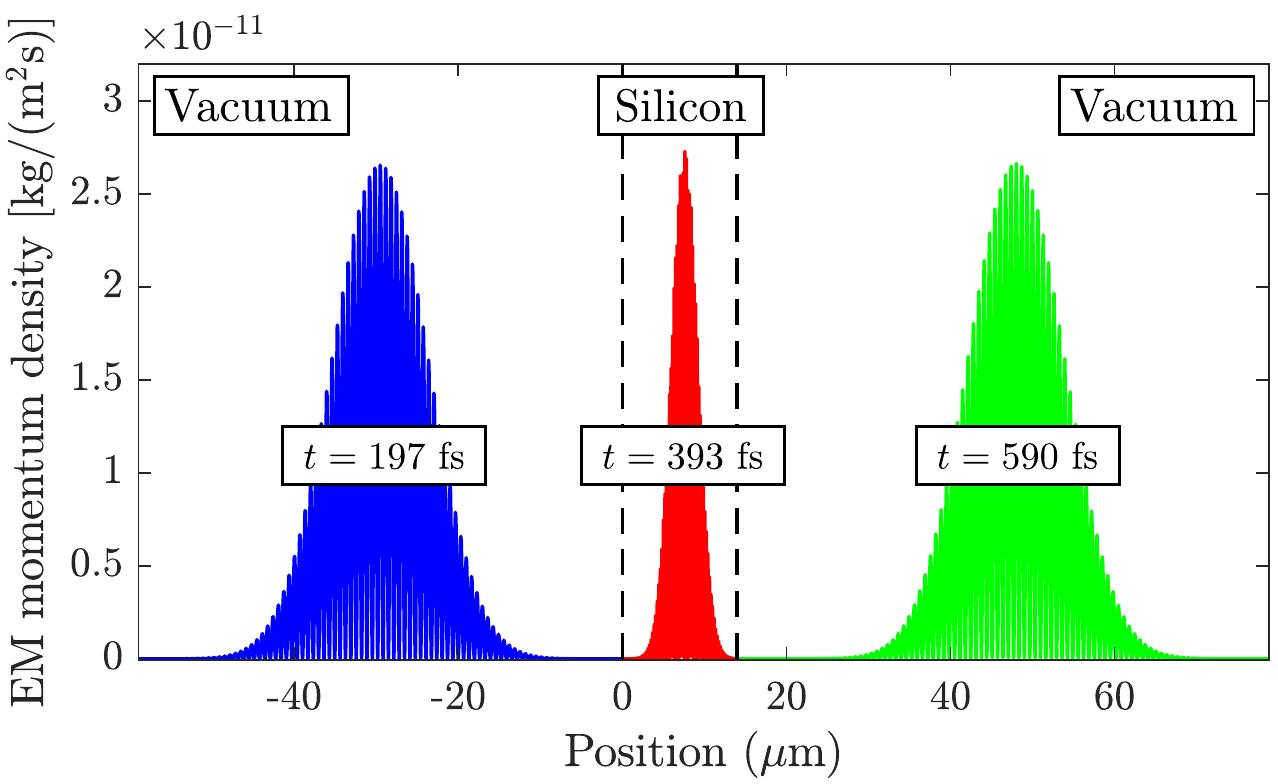}
\vspace{-0.45cm}
\caption{\label{fig:propagation}
Position dependence of the EM momentum density at three instances of time during the propagation of a Gaussian light pulse through a silicon crystal. First, the field is incident from the left. Second, the field is inside the crystal. Third, the field has left the crystal. The vertical dashed lines show the positions of the two interfaces of the crystal. The momentum densities reflected from the first and second interfaces are very small and not shown in the figure.}
\vspace{-0.2cm}
\end{figure}

The propagation of the Gaussian light pulse through the silicon crystal is illustrated in Fig.~\ref{fig:propagation}, where the EM momentum density is plotted at three instances of time: \emph{first}, before the crystal ($t=197$ fs), \emph{second}, inside the crystal ($t=393$ fs), and \emph{third}, after the crystal ($t=590$ fs). The very small reflections from the first and the second interface are not visible because of their smallness. The EM energy reflected from the first interface is 0.14\% of the total incident EM energy, and the EM energy reflected from the second interface is also 0.14\% of the total incident EM energy. Thus, the total EM energy lost from the incident pulse in the full transmission process is 0.28\%. One can conclude that the antireflection property of the interfaces is relatively well optimized. Inside the crystal the pulse in Fig.~\ref{fig:propagation} is spatially much narrower than in vacuum. This is due to the lower group velocity of light in the crystal compared to that in vacuum. We can also see that the largest magnitudes of the EM momentum density in the crystal and in vacuum are equal. This follows from the continuity of the Poynting vector.

The total EM momentum of the pulse is the volume integral of the EM momentum density. Neglecting the small reflection, the ratio of these integrals in the crystal and in vacuum is equal to the ratio of the corresponding group velocities. Thus, the EM momentum in the crystal is $p_\mathrm{EM}=p_0/n_\mathrm{g}$, where $p_0$ is the momentum of the incident EM field. The momentum of the atomic MDW is equal to $p_\mathrm{MDW}=(n_\mathrm{p}-1/n_\mathrm{g})p_0$. Thus, the total momentum of the coupled MP state of light in the crystal becomes $p_\mathrm{MP}=n_\mathrm{p}p_0$. This phase index proportionality of the total momentum of light is in agreement with the high-precision measurements of radiation pressure in liquids by Jones \emph{et al.}~\cite{Jones1978} and the measurements of the recoil momentum of atoms in a dilute Bose-Einstein condensate gas by Campbell \emph{et al.}~\cite{Campbell2005}.

\section{\label{sec:SEM}SEM tensors and their relativistic invariance}

The SEM tensor of a physical system collects the energy and momentum densities and the stress tensor in a single physical quantity. The general contravariant form of an arbitrary SEM tensor in the Minkowski space-time is defined by
$\mathbf{T}=T^{\alpha\beta}\mathbf{e}_\alpha\otimes\mathbf{e}_\beta$, where the Einstein summation convention is used, and $\mathbf{e}_\alpha$ and $\mathbf{e}_\beta$ are unit vectors of the space-time. The Greek indices range over the four components of the space-time, i.e., $(ct,x,y,z)$. The corresponding matrix representation of $\mathbf{T}$ is given by \cite{Landau1989,Jackson1999,Misner1973}
\begin{equation}
 \mathbf{T}=
 \left[\begin{array}{cc}
  W & c\mathbf{G}^T\\
  c\mathbf{G} & \boldsymbol{\mathcal{T}}\\
 \end{array}\right]
 =\left[\begin{array}{cccc}
  W & cG^x & cG^y & cG^z\\
  cG^x & \mathcal{T}^{xx} & \mathcal{T}^{xy} & \mathcal{T}^{xz}\\
  cG^y & \mathcal{T}^{yx} & \mathcal{T}^{yy} & \mathcal{T}^{yz}\\
  cG^z & \mathcal{T}^{zx} & \mathcal{T}^{zy} & \mathcal{T}^{zz}
 \end{array}\right].
 \label{eq:emt}
\end{equation}
Here the superscript $T$ denotes the transpose. Since the SEM tensor of the material, made from the quantities discussed in Sec.~\ref{sec:separation}, is symmetric by definition, so is also the EM SEM tensor since the total SEM tensor of the system must be symmetric. Therefore, all of our SEM tensors are symmetric and strictly based on the classical definition in Eq.~\eqref{eq:emt}.

The SEM tensor $\mathbf{T}_\mathrm{mat}$ of the material can be formed by substituting the energy density, momentum density, and the stress tensor of the material from Sec.~\ref{sec:separation} into Eq.~\eqref{eq:emt}. Even more compactly, $\mathbf{T}_\mathrm{mat}$ can be written using the four-velocity of the material, given by $\mathbf{U}_\mathrm{a}=\gamma_{\mathbf{v}_\mathrm{a}}(c,\mathbf{v}_\mathrm{a})$. The resulting compact form of $\mathbf{T}_\mathrm{mat}$ is given by \cite{Dirac1996,Misner1973}
\begin{equation}
 \mathbf{T}_\mathrm{mat}
=\frac{\rho_\mathrm{a}}{\gamma_{\mathbf{v}_\mathrm{a}}^2}\mathbf{U}_\mathrm{a}\otimes\mathbf{U}_\mathrm{a}.
\label{eq:MATSEMG}
\end{equation}
Consequently, the SEM tensor of the field is also unambiguously determined in all inertial frames. Note that accounting for the elastic force density and the electro- and magnetostriction will modify the SEM tensor of the material in Eq.~\eqref{eq:MATSEMG}. The resulting modified SEM tensor of the material is expected to be of the form presented in Ref.~\cite{Hernandez1970}. Its detailed dynamical study under the influence of the optical field is left as a topic of a future work.

Following the approach presented in Sec.~IVB of Ref.~\cite{Partanen2021b} to use the Lorentz transformation to determine the SEM tensor of the field in an arbitrary inertial frame once the SEM tensor of the field in the laboratory frame is known, in the present case, we obtain
\begin{align}
 \mathbf{T}_\mathrm{EM}
 &=\frac{1}{2}(\mathbf{F}\boldsymbol{g}\boldsymbol{\mathcal{D}}+\boldsymbol{\mathcal{D}}\boldsymbol{g}\mathbf{F})
 -\frac{1}{4}\boldsymbol{g}\mathrm{Tr}(\mathbf{F}\boldsymbol{g}\boldsymbol{\mathcal{D}}\boldsymbol{g})\nonumber\\
 &\hspace{0.5cm}-\frac{1}{2c^2}\Big[(\mathbf{F}\boldsymbol{g}\boldsymbol{\mathcal{D}}-\boldsymbol{\mathcal{D}}\boldsymbol{g}\mathbf{F})\boldsymbol{g}(\mathbf{U}_\mathrm{a0}\otimes\mathbf{U}_\mathrm{a0})\nonumber\\
 &\hspace{0.5cm}+(\mathbf{U}_\mathrm{a0}\otimes\mathbf{U}_\mathrm{a0})\boldsymbol{g}(\boldsymbol{\mathcal{D}}\boldsymbol{g}\mathbf{F}-\mathbf{F}\boldsymbol{g}\boldsymbol{\mathcal{D}})\Big]\nonumber\\
 &\hspace{0.5cm}+\frac{\rho_\mathrm{EM,disp}}{\gamma_{\mathbf{v}_\mathrm{a0}}^2}\mathbf{U}_\mathrm{a0}\otimes\mathbf{U}_\mathrm{a0}.
 \label{eq:TAG}
\end{align}
Here $\mathbf{U}_\mathrm{a0}=\gamma_{\mathbf{v}_\mathrm{a0}}(c,\mathbf{v}_\mathrm{a0})$ is the four-velocity of the rest frame of the material, $\mathrm{Tr}(x)$ is the trace of a matrix, and $\boldsymbol{g}=g_{\alpha\beta}\mathbf{e}^\alpha\otimes\mathbf{e}^\beta$, with $g_{00}=1$, $g_{11}=g_{22}=g_{33}=-1$, is the diagonal matrix representation of the Minkowski metric tensor. The EM field tensor $\mathbf{F}$ and the EM displacement tensor $\boldsymbol{\mathcal{D}}$ in Eq.~\eqref{eq:TAG} are given in contravariant forms $\mathbf{F}=F^{\alpha\beta}\mathbf{e}_\alpha\otimes\mathbf{e}_\beta$ and $\boldsymbol{\mathcal{D}}=\mathcal{D}^{\alpha\beta}\mathbf{e}_\alpha\otimes\mathbf{e}_\beta$ as
\begin{align}
&\mathbf{F}=\left[\begin{array}{cccc}
0 & -E_x/c & -E_y/c & -E_z/c\\
E_x/c & 0 & -B_z & B_y\\
E_y/c & B_z & 0 & -B_x\\
E_z/c & -B_y & B_x & 0
\end{array}\right],\nonumber\\
&\hspace{0.2cm}\boldsymbol{\mathcal{D}}=\left[\begin{array}{cccc}
0 & -D_xc & -D_yc & -D_zc\\
D_xc & 0 & -H_z & H_y\\
D_yc & H_z & 0 & -H_x\\
D_zc & -H_y & H_x & 0
\end{array}\right]\!.
\label{eq:FDtensors}
\end{align}
The term $\rho_\mathrm{EM,disp}$ appearing in Eq.~\eqref{eq:TAG} is an effective EM mass density term associated with dispersion. In the special case of the laboratory frame, it is defined through the last two terms of the EM energy density in Eq.~\eqref{eq:WEM} as
\begin{equation}
 \rho_\mathrm{EM,disp}^\mathrm{(L)}=\frac{\omega_0}{2c^2}\Big(\frac{\partial\varepsilon}{\partial\omega_0}\langle|\mathbf{E}^\mathrm{(L)}|^2\rangle+\frac{\partial\mu}{\partial\omega_0}\langle|\mathbf{H}^\mathrm{(L)}|^2\rangle\Big).
\end{equation}
This quantity transforms between inertial frames as a mass density and it is very convenient in expressing the relativistically invariant SEM tensor of the EM field in Eq.~\eqref{eq:TAG}.

The SEM tensor $\mathbf{T}_\mathrm{mat,0}$ of the material in the absence of light is given by a similar form as the instantaneous SEM tensor $\mathbf{T}_\mathrm{mat}$ of the material under the influence of the EM field in Eq.~\eqref{eq:MATSEMG}. Thus, $\mathbf{T}_\mathrm{mat,0}$ is given by
\begin{equation}
 \mathbf{T}_\mathrm{mat,0}=\frac{\rho_\mathrm{a0}}{\gamma_{\mathbf{v}_\mathrm{a0}}^2}\mathbf{U}_\mathrm{a0}\otimes\mathbf{U}_\mathrm{a0}.
\label{eq:MAT0SEMG}
\end{equation}
The SEM tensor of the atomic MDW is defined as the difference $\mathbf{T}_\mathrm{MDW}=\mathbf{T}_\mathrm{mat}-\mathbf{T}_\mathrm{mat,0}$. Substituting the expressions of the tensors $\mathbf{T}_\mathrm{mat}$ and $\mathbf{T}_\mathrm{mat,0}$ into this difference then gives
\begin{equation}
 \mathbf{T}_\mathrm{MDW}=\frac{\rho_\mathrm{a}}{\gamma_{\mathbf{v}_\mathrm{a}}^2}\mathbf{U}_\mathrm{a}\otimes\mathbf{U}_\mathrm{a}
-\frac{\rho_\mathrm{a0}}{\gamma_{\mathbf{v}_\mathrm{a0}}^2}\mathbf{U}_\mathrm{a0}\otimes\mathbf{U}_\mathrm{a0}.
\label{eq:MDWSEMG}
\end{equation}
Being given in terms of the EM field and displacement tensors, four-velocities, and Lorentz scalar factors, all SEM tensors presented in this section are transparently covariant, i.e., they have the same expressions in all inertial frames, thus, satisfying the conditions of relativistic invariance.

Note that the sum of the last three terms of $\mathbf{T}_\mathrm{EM}$ in Eq.~\eqref{eq:TAG} can be viewed as a relativistically consistent generalization of the Abraham SEM tensor, where the Abraham SEM tensor is allowed to depend on $\mathbf{U}_\mathrm{a0}$. This has been discussed in previous literature, see, e.g., Eq.~(42) of Ref.~\cite{Obukhov2008}, Eq.~(2.10) of Ref.~\cite{Makarov2011}, and Eq.~(41) of Ref.~\cite{Partanen2021b}. Consequently, \emph{for this generalization of the Abraham SEM tensor}, the previous claim \cite{Veselago2009,Veselago2010,Wang2013a,Sheppard2016a,Kemp2017} that the Abraham SEM tensor would be relativistically invalid is no longer true.

\section{\label{sec:conclusions}Conclusions}

In conclusion, we have presented an unambiguous position- and time-dependent optical force theory applicable to simulations of the propagation of light pulses in inhomogeneous dispersive materials. With existing simulation tools, the theory enables detailed modeling of the optomechanics of 3D photonic materials and devices. For example, the present theory has been applied to negative-index metamaterials in Ref.~\cite{Partanen2022a}. For absorbing materials, the well-known Lorentz force density should be added to the force density of the present work to describe also forces related to absorption. In the nonlinear optics regime, other phenomena, such as the Kerr effect, contribute and, accordingly, one must separately consider effects like electrostriction. The present classical field theory must be extended to the quantum domain to describe also the local torque originating from the interaction of the spin of light with the atoms.

\begin{acknowledgments}
This work has been funded by the Academy of Finland under Contract No.~318197 and 349971 and European Union's Horizon 2020 Marie Sk\l{}odowska-Curie Actions (MSCA) individual fellowship under Contract No.~846218. Aalto Science-IT is acknowledged for computational resources.
\end{acknowledgments}

\end{document}